\documentclass[11pt,twoside]{article}
\usepackage{asp2014}

\aspSuppressVolSlug
\resetcounters

\begin{document}

\title{Cool White Dwarfs Selection with Pan-STARRS Proper Motions}
\author{M. C. Lam and N. C. Hambly}
\affil{Institute for Astronomy, University of Edinburgh, Royal Observatory, Blackford Hill, Edinburgh EH9 3HJ, UK}

\paperauthor{M. C. Lam}{mlam@roe.ac.uk}{}{Institute for Astronomy}{University of Edinburgh}{Royal Observatory, Blackford Hill}{Edinburgh}{EH9 3HJ}{UK}
\paperauthor{N. C. Hambly}{nch@roe.ac.uk}{}{Scottish Universities Physics Alliance (SUPA), Institute for Astronomy}{University of Edinburgh}{Royal Observatory, Blackford Hill}{Edinburgh}{EH9 3HJ}{UK}

\begin{abstract}
The use of Reduced Proper Motion in identifying isolated white dwarfs has long been used as a proxy for the absolute magnitude in a population with known kinematics. This, however, introduces a proper motion detection limit on top of the existing photometric limit. How the survey volume is hampered by this extra parameter is discussed in \cite{Hambly2012}. In this work, we discuss some robust outlier rejection methods in order to minimise the proper motion limit and hence maximise the survey volume. The generalised volume, corrected for the distance of the Sun from the Galactic Plane, is integrated explicitly.
\end{abstract}

\section{Introduction}
Main sequence (MS) stars with masses of less than ~8 M$_{\odot}$ will end up as white dwarfs (WDs). This mass range encompasses the vast majority of stars in the Galaxy. Thus a WD, which is a degenerate core left behind from its progenitor, is the most common end-point of stellar evolution. Nuclear burning is negligible at this stage, so WDs cannot replenish the energy they radiate away. Hence, the luminosity and temperature decrease monotonically with time. The WD luminosity function (LF) was first used as a cosmochronometer half a century ago. Given a finite age of the Galaxy, there is a temperature beyond which the oldest WDs have not had time to reach which would translate to a sudden downturn in the WDLF. Further to cosmochronometry, a WDLF can be inverted to provide the star formation history\,\citep{Rowell2013}.

\section{Pan-STARRS}
Pan-STARRS-1 (PS1) is a 1.8\,m optical wide-field imager developed by the University of Hawaii (See \cite{Metcalfe2013} and references therein for details). The PS1 3$\pi$ Steradian Survey covers the sky north of declination $-30^{\circ}$ with five broadband filters designated as g$_{\text{p1}}$, r$_{\text{p1}}$, i$_{\text{p1}}$, z$_{\text{p1}}$ and y$_{\text{p1}}$ spanning 400-1000\,nm. The sky has been imaged 60 times on average in the 4-year survey, allowing PS1 to determine accurate proper motions with small epoch differences. The test data covers the sky between declination $0^{\circ}$ and $7.5^{\circ}$. The selection criteria are:

\begin{enumerate}
\item A minimum of 1.5 years epoch difference
\item Proper motion with 10$\sigma$ confidence
\item Detection in all five filters
\item Good morphology flags
\end{enumerate}

\subsection{Proper Motions}
In the PS1 reduction pipeline, the proper motions and the parallaxes are calculated simultaneously for all objects having sufficient coverage in parallax factor. However, most objects would not have detectable parallaxes such that some objects can have either spurious proper motions or unrepresentative uncertainties and $\chi^{2}$ values for the proper motions. Furthermore, in a proper motion limited sample, the sampling volume scales between $\mu^{2}$ and $\mu^{3}$ depending on the population \citep{Hambly2012}. Therefore, a smaller lower proper motion limit would greatly increase the survey volume. We have investigated some robust algorithms by rejecting outliers to improve the quality of proper motion, hence to reduce the proper motion limit to be adopted:

\begin{enumerate}
\item \textbf{Fitting by minimising absolute deviation}\\
With this method, data points are unit weighted. It is found to be ineffective at handling clustered data.
\item \textbf{Jackknife method}\\
This would be the ideal option, however, there is not enough computing power to go through this process (jackknifing 10 points out of the 60 epoch measurements for $10^{9}$ objects would require $10^{26}$ calculations).
\item \textbf{Iterative outlier rejection}\\
This method identifies data points lying outside 3$\sigma$ from the best fit solution as outliers. A new weighted-least square solution will be found and this process continues until no more data points are rejected. This method is much faster than jackknife, but it is more sensitive to data further away from the centroid of the solution.
\item \textbf{Improved iterative outlier rejection}\\
The improved version has the tolerance level based on the propagation of errors which accepts larger deviations when the data points are further away from the centroid.
\end{enumerate}
The best fit line can be described by the function
\begin{equation}
f = x + \mu \, t
\end{equation}
and the uncertainty is given by the standard propagation of errors using the covariance matrix $\left\langle \sigma_{i} \, \sigma_{j} \right\rangle$
\begin{align}
\begin{split}
\sigma^{2} &= \sum_{i} \sum_{j} \left\langle \sigma_{i} \, \sigma_{j} \right\rangle \frac{\partial f}{\partial \rho_{i}} \frac{\partial f}{\partial \rho_{j}} \\
&= \sigma^{2}_{x} \frac{\partial f}{\partial x}^{2} + 2 \sigma^{2}_{x \mu} \frac{\partial f}{\partial x} \frac{\partial f}{\partial \mu} + \sigma^{2}_{\mu} \frac{\partial f}{\partial \mu}^{2} \\
&= \sigma^{2}_{x} + \sigma^{2}_{\mu} t^{2}
\end{split}
\end{align}

\section{Methods}
\subsection{Model Atmosphere}
The synthetic colour of both DA, DB and mixed hydrogen-helium model atmospheres in the Pan-STARRS colours are provided by Dr. Bergeron\,(\citealt{Holberg2006}; \citealt{Kowalski2006}; \citealt{Tremblay2011}; \citealt{Bergeron2011}) which were based on the most recent calibrations. At this early stage of analysis, it is assumed that all WDs have pure hydrogen atmosphere and have surface gravity $\text{log}(g) = 8.0$ in order to fit the effective temperatures and the distances simultaneously.

\subsection{Reduced Proper Motion}
Since WDs have much smaller radii than MS stars at the same temperature, the WD cooling sequence is a few magnitudes fainter than the main sequence. An efficient way to select WD candidates is to use reduced proper motion\,(H)\,(\citealt{Kilic2006}; \citealt{Harris2006}). By using proper motions as proxy-parallaxes, one can obtain the reduced proper motion, which is analogous to the absolute magnitude:
\begin{equation}
\text{H}_{\text{r}_{\text{p1}}} = 5 + 5\,\text{log}(\mu) + \text{r}_{\text{p1}}
\end{equation}
Using proper motion and the photometric parallax, the tangential velocities, v$_{\text{tan}}$, can be deduced. This is an important quantity in deriving the distance limit due to the proper motion limit. By holding v$_{\text{tan}}$ constant, one can determine how far the object can be placed before its proper motion would drop below the lower proper motion limit.

\subsection{Generalised Schmidt Estimator}
In the $\frac{1}{\text{V}_{\text{max}}}$ method, the contribution of each object to the LF is weighted by the inverse of the maximum volume in which the object could be observed by the survey. However, this technique assumes objects are uniformly distributed. In reality, stars in the solar neighbourhood are concentrated in the plane of the disk so the effects of space-density gradient have to be corrected. This led to the development of the generalised volume $\text{V}_{\text{gen}}$\,\citep{Stobie1989} which is calculated by integrating the appropriate stellar density profile $\frac{\rho(\text{r})}{\rho_{\odot}} = \exp(-\frac{|\text{z}|}{\text{H}}) = \exp(-\frac{|\text{r}\sin(\text{b})|}{H})$ along the line of sight,
\begin{equation}
\text{V}_{\text{gen}} = \chi(\text{v}_{\text{tan}}) \, \Omega \int_{\text{d}_{\text{min}}}^{\text{d}_{\text{max}}} \frac{\rho(\text{r})}{\rho_{\odot}} \, \text{r}^{2} \, \text{dr}
\end{equation}
where $\chi(\text{v}_{\text{tan}})$ is the discovery fraction of the sample from the lower tangential velocity limit, $\Omega$ is the size of the solid angle of the survey, d$_{\text{min}}$ and d$_{\text{max}}$ are the distances limits set by the bright and faint detection limits as well as the high and low proper motion limits, H is the scale height of the Galactic disk profile. By further taking into account of the distance of the Sun from the Galactic Plane, the density profile becomes $\frac{\rho(\text{r})}{\rho_{\odot}} = \exp\left(-\frac{|\text{r}\sin(\text{b})+\text{z}_{\odot}|}{\text{H}}\right)$ and the integral can be solved analytically:
\begin{equation}
\text{V}_{\text{gen}} =
\begin{cases}
-\exp\left({-\frac{\text{d}_{\text{max}}\sin(\text{b})+\text{z}_{\odot}}{\text{H}}}\right)\left(2\xi^{3} + 2\text{d}_{\text{max}}\xi^{2}+\text{d}_{\text{max}}^{2}\xi\right), & \text{if} \left[ \text{d}_{\text{max}}\sin(\text{b})+\text{z}_{\odot} \right] \geq 0\\
\exp\left({\frac{\text{d}_{\text{max}}\sin(\text{b})+\text{z}_{\odot}}{\text{H}}}\right)\left(2\xi^{3} - 2\text{d}_{\text{max}}\xi^{2}+\text{d}_{\text{max}}^{2}\xi\right), & \text{otherwise},
\end{cases}
\end{equation}
where $\xi = \left( \text{H}/\sin{\left(\text{b}\right)} \right)$ .
\section{Future Work}
The final data release of PS1 is scheduled to be in mid 2015. The data quality is expected to increase with the improvement in the reduction pipeline and the increase in the maximum epoch difference. The total number of WD candidates with v$_{\text{tan}} > 40\,\text{km}\,\text{s}^{-1}$ is expected to be about 40000 \citep{Hambly2012}. With sufficient objects, it is possible to untangle the thin disk, thick disk and stellar halo which did not arrive at statistically confident results in previous work \citep{Rowell2011}. Furthermore, the significant increase in the survey volume allows the inversion of the luminosity function to recover the star formation history of the Galaxy with greater precision.

\hfill
\acknowledgements The Pan-STARRS1 Surveys (PS1) have been made possible through contributions of the Institute for Astronomy, the University of Hawaii, the Pan-STARRS Project Office, the Max-Planck Society and its participating institutes, the Max Planck Institute for Astronomy, Heidelberg and the Max Planck Institute for Extraterrestrial Physics, Garching, The Johns Hopkins University, Durham University, the University of Edinburgh, Queen's University Belfast, the Harvard-Smithsonian Center for Astrophysics, the Las Cumbres Observatory Global Telescope Network Incorporated, the National Central University of Taiwan, the Space Telescope Science Institute, the National Aeronautics and Space Administration under Grant No. NNX08AR22G issued through the Planetary Science Division of the NASA Science Mission Directorate, the National Science Foundation under Grant No. AST-1238877, the University of Maryland, and Eotvos Lorand University (ELTE).


\begin{thebibliography}{}

\bibitem[Bergeron et al.(2011)]{Bergeron2011}
Bergeron, P., Wesemael, F., Dufour, P., Beauchamp, A., Hunter, C., Saffer, R. A., Gianninas, A., Ruiz, M. T., Limoges, M.-M., Dufour, P., Fontaine, G., Liebert, J., 2011, ApJ, 737, 28

\bibitem[Hambly et al.(2012)]{Hambly2012}
Hambly, N. C., Rowell, N., Tonry, J. L., Magnier, E. A. and Stubbs, C. W. 2012, ASPCS, 469, 253

\bibitem[Harris et al.(2006)]{Harris2006}
Harris, H. C., Munn, J. A., Kilic, M., Liebert, J., Williams, K. A., von Hippel, T., Levine, S. E., Monet, D. G.; Eisenstein, D. J., Kleinman, S. J., Metcalfe, T. S., Nitta, A., Winget, D. E., Brinkmann, J., Fukugita, M., Knapp, G. R., Lupton, R. H., Smith, J. A., Schneider, D. P. 2006, AJ, 131, 571

\bibitem[Holberg \& Bergeron(2006)]{Holberg2006}
Holberg, J. B., Bergeron, P., 2006, AJ, 132, 1221

\bibitem[Kilic et al.(2006)]{Kilic2006}
Kilic, M., Munn, J. A., Harris, H. C., Liebert, J., von Hippel, T., Williams, K. A., Metcalfe, T. S., Winget, D. E., Levine, S. E., 2006, AJ, 131, 582

\bibitem[Kowalski \& Saumon(2006)]{Kowalski2006}
Kowalski, P. M., Saumon, D., 2006, ApJ, 651, L137

\bibitem[Metcalfe et al.(2013)]{Metcalfe2013}
Metcalfe, N., Farrow, D. J., Cole, S., Draper, P. W., Norberg, P., Burgett, W. S., Chambers, K. C., Denneau, L., Flewelling, H., Kaiser, N., Kudritzki, R., Magnier, E. A., Morgan, J. S., Price, P. A., Sweeney, W., Tonry, J. L., Wainscoat, R. J., Waters, C., 2013, MNRAS, 435, 1825

\bibitem[Rowell \& Hambly(2011)]{Rowell2011}
Rowell, N., Hambly, N. C. 2011, MNRAS, 417, 93

\bibitem[Rowell(2013)]{Rowell2013}
Rowell, N., 2013, MNRAS, 434, 1549

\bibitem[Stobie, Ishida \& Peacock(1989)]{Stobie1989}
Stobie, R. S., Ishida, K., Peacock, J. A., 1989, MNRAS, 238, 709

\bibitem[Tremblay, Bergeron \& Gianninas(2011)]{Tremblay2011}
Tremblay, P.-E., Bergeron, P., Gianninas, A., 2011, ApJ, 730, 128

\end{thebibliography}
\end{document}